\documentclass[twocolumn, showkeys, showpacs, preprintnumbers, prd, superscriptaddress, nofootinbib, shortlabels]{revtex4-1}
\usepackage[T1]{fontenc}       
\usepackage[utf8]{inputenc}    
\usepackage[english]{babel}    
\usepackage{amsmath, amssymb, amsfonts, mathtools, bm}
\usepackage{booktabs}          
\usepackage{multirow}          
\usepackage{dcolumn}           
\usepackage[table]{xcolor}     
\usepackage{graphicx}
\usepackage{epstopdf}        
\usepackage{soul}                       
\usepackage{enumitem}          
\usepackage{pifont}            
\usepackage{siunitx}           
\usepackage{braket}           
\usepackage{placeins}
\usepackage{afterpage}
\usepackage{transparent}       
\usepackage{hyperref}         
\usepackage{geometry}
\begin{document}

\title{Observational constraints on the product of dark energy chemical potential and number density in out-of-equilibrium  models}

\author{J. M. Costa Netto}
\email{jose.medeiroscn@gmail.com}
\affiliation{Núcleo de Formação Docente, Centro Acadêmico do Agreste, Universidade Federal de Pernambuco, Caruaru, Pernambuco, 55014-900, Brazil}

\author{Javier E. Gonzalez}
\email{javiergonzalezs@academico.ufs.br}
\affiliation{Departamento de Física, Universidade Federal de Sergipe, São Cristóvão, Sergipe, 49107-230, Brazil}

\author{H. H. B. Silva}
\email{heydson.henrique@ufpe.br} 
\affiliation{Núcleo de Formação Docente, Centro Acadêmico do Agreste, Universidade Federal de Pernambuco, Caruaru, Pernambuco, 55014-900, Brazil}
\begin{abstract}
    In this work, we impose observational limits on the product of dark energy chemical potential, $\mu$, and number density, $n$, at the present time in out-of-equilibrium models, considering that particles can be created or destroyed in the fluid at a rate $\Gamma=3\alpha H(a)$, where $\alpha$ is a constant and $H(a)\equiv\dot{a}/a$ is the Hubble parameter. We combine the bounds derived from the positivity of entropy and the second law of thermodynamics with observational constraints on the Chevallier-Polarski-Linder (CPL) and Barboza-Alcaniz (BA) parameterizations of the equation of state (EoS) of the component. We use Type Ia supernovae (SN Ia) data from Pantheon+; baryon acoustic oscillation (BAO) data from DESI DR2; and cosmic microwave background (CMB) measurements from Planck. For $\alpha>0$ (particle creation), the thermodynamic restrictions yield only upper limits for the $\mu_{0}n_{0}$ product, while in the case of $\alpha<0$ (particle destruction) they establish both  upper and lower limits, allowing for a range of values to be obtained. In both scenarios, however, we find that the chemical potential of dark energy must be negative, $\mu<0$, which indicates a preference for the phantom regime. In particular, when $\alpha<0$, it is noted that the thermodynamic bounds are simultaneously compatible only for very small absolute values of $\alpha$, with $\alpha=-0.0002$ being the limiting case and resulting in $\mu_{0}n_{0}(\alpha=-0.0002)=-2.2_{-0.7}^{+1.0}\,\,GeV/m^{3}$.
\end{abstract}
\maketitle

\section{Introduction} \label{sec1}

Recent results obtained by the Dark Energy Spectroscopic Instrument (DESI) \cite{desi1, desi2} indicate, when combining its baryon acoustic oscillations (BAO) data with Type Ia supernovae (SN Ia) data \cite{pantheon2022.1, pantheon2022.2, union3, des} and cosmic microwave background (CMB) measurements \cite{planck2018.1, planck2018.2} from other collaborations, a preference toward dynamic dark energy models, more specifically models in which the equation of state (EoS) parameter of the component is a function of the scale factor $a$, $w=w(a)$. There is also a preference for a transition between a phantom regime and quintessence scenarios, where $w(a)<-1$ and $|w(a)|\leq 1$, respectively. Although these conclusions are new and subject of much debate given their impact, the study of dark energy and, in particular, dynamical models, hence, becomes a topic of great relevance in the world of cosmology.

Dark energy can be studied, essentially, from two perspectives: one in which it is modeled by a relativistic fluid and the entire basis of thermodynamics is applied \cite{lima2004, lima2008, pereira, silva2012, silva2013,Gonzalez:2018rop}; another in which it is associated with a scalar potential and field theory is used \cite{ratra, caldwell, zlatev, peebles}. Although the two approaches are not incompatible, in this work we adopt the thermodynamic one. For models in which the barotropic EoS parameter

\begin{equation} \label{eq1}
    w(a)\equiv\frac{p_{x}}{\rho_{x}}=w_{0}+w_{a}f(a),
\end{equation}

\noindent where $p_{x}$ and $\rho_{x}$ are, respectively, the pressure and the dark energy density, $w_{0}$ and $w_{a}$ are constants, and $f(a)$ is a scale factor function (or, equivalently,  time dependent), the fluid departs from the adiabatic regime and effectively behaves as if it possessed a bulk viscous pressure, generating entropy as an \textit{a posteriori} effect \cite{silva2012, silva2013}. Independently of whether dark energy is in or out of equilibrium, the chemical potential, $\mu$, is responsible for determining the range of possible values of $w(a)$ \cite{silva2013, lima2008, pereira}: if $\mu<0$, the component will be in the phantom regime; if $\mu=0$, the cosmological constant is recovered; if $\mu>0$, neither of these two cases is reached.

In cosmology, it is well known that bulk viscosity is closely related to particle creation or destruction processes. Indeed, it is possible to phenomenologically describe bulk viscous pressures through particle sources or sinks \cite{calvao, lima1992, zimdahl1993, zimdahl1996.2, lima1996}. Bearing in mind that dark energy begins to mimic a bulk viscous pressure as a consequence of a variable EoS, it is reasonable to assume the possibility of particles being created or destroyed in the fluid. In \cite{costanetto} we consider exactly that by proposing

\begin{equation} \label{eq2}
    N^{\mu}_{\hspace{1ex};\mu}=n\Gamma,
\end{equation}

\noindent where $N^{\mu}$ is the particle current vector, $n$ is the number density, and $\Gamma$ is the rate of particle creation $(\Gamma>0)$ or destruction ($\Gamma<0$). This, of course, adds a new term to the entropy source and, therefore, affects not only the evolution of $n$, but also the entropy density $s$. However, since the energy-momentum tensor $T^{\mu\nu}$ is unaffected, the continuity equation and thus the evolution of the dark energy density remains the same --- as does the temperature evolution law, since it depends only on the EoS.

Although we perform an initial generalization of the out-of-equilibrium thermodynamic models of dark energy, we did not specify a functional form for the $\Gamma$-term in \cite{costanetto}. Based on \cite{lima1996}, in this present work we adopt the expression  

\begin{equation} \label{eq3}
    \Gamma=3\alpha H(a),
\end{equation}

\noindent where $\alpha$ is assumed to be constant and can be positive (particle creation) or negative (particle destruction), and $H(a)\equiv\dot{a}/a$ is the Hubble parameter. Our objective is to impose observational constraints on $\mu_{0}n_{0}$, the present time value of the product of the chemical potential and the number density, using the bounds derived from the positivity of entropy and the second law of thermodynamics.

This work is organized as follows: in Section \ref{sec2}, we review the model proposed in \cite{costanetto}. Starting from the fundamental equations that describe the fluid, we arrive at expressions for the entropy and the entropy source that allow us to obtain the thermodynamic bounds mentioned above. We combine these limits with observational constraints on the Chevallier-Polarski-Linder (CPL) \cite{chevallier, linder} and Barboza-Alcaniz (BA) \cite{barboza} parameterizations of the barotropic EoS parameter of dark energy. To this end, we rely on SN Ia data from Phanteon+ \cite{pantheon2022.1, pantheon2022.2}; BAO data from DESI DR2 \cite{desi2}; and CMB measurements from Planck \cite{planck2018.1, planck2018.2}. These datasets and its respective statistical treatment are contained in Section \ref{sec3} (and in Appendix \ref{ap:BA}). The limits imposed on the product of chemical potential and particle density are found in Section \ref{sec4} (and in Appendix \ref{ap:BA}), along with a detailed discussion of their implications. In Section \ref{concl}, we present the conclusions of the research.

Throughout this work, we use the subscript $0$ to indicate quantities measured at the present time, $t=t_{0}$, and, unless otherwise stated, we adopt units in which $8\pi G=c=1$ and $a(t_{0})=a_{0}=1$. Our metric signature is $(+,-,-,-)$.
\section{Out-of-equilibrium dark energy thermodynamics} \label{sec2}

When working with relativistic hydrodynamics, the basic tensors of the dark energy fluid in a flat Friedmann-Lemaître-Robertson-Walker (FLRW) universe are \cite{weinberg1972, weinberg2008}:

\begin{equation} \label{eq4}
    T^{\mu\nu}=\rho_{x}U^{\mu}U^{\nu}-p_{x}h^{\mu\nu},
\end{equation}

\begin{equation} \label{eq5}
    N^{\mu}=nU^{\mu},
\end{equation}

\begin{equation} \label{eq6}
    S^{\mu}=sU^{\mu}=n\sigma U^{\mu},
\end{equation}

\noindent where $U^{\mu}$ is the 4-velocity, $h^{\mu\nu}\equiv g^{\mu\nu}-U^{\mu}U^{\nu}$ is the projector onto the local rest space of the 4-velocity, $S^{\mu}$ is the entropy current vector, and $\sigma$ is the specific entropy. 

If the fluid is outside the adiabatic limit, we introduce the variations $\Delta T^{\mu\nu}$ and $\Delta N^{\mu}$ in Eq. (\ref{eq4}) and Eq. (\ref{eq5}), respectively \cite{silva2002}:

\begin{equation} \label{eq7}
    T^{\mu\nu}=\rho_{x}U^{\mu}U^{\nu}-p_{x}h^{\mu\nu}+\Delta T^{\mu\nu},
\end{equation}

\begin{equation} \label{eq8}
    N^{\mu}=nU^{\mu}+\Delta N^{\mu}.
\end{equation}

\noindent The Eckart \cite{eckart} and Landau-Lifshitz \cite{landau} approaches are the most common in cosmology when it comes to non-equilibrium thermodynamics. Although the former defines $U^{\mu}$ as the 4-velocity of particle transport and the latter defines it as the 4-velocity of energy transport, both formalisms consider that the out-of-equilibrium terms present in Eqs. (\ref{eq7}, \ref{eq8}) are small enough to consider only first-order deviations. Assuming that scalar processes are the only dissipative effects occurring in dark energy, it can be shown that \cite{silva2002}

\begin{equation} \label{eq9}
    U_{\mu}\Delta T^{\mu\nu}_{\hspace{1.5ex};\nu}=(\Pi+p_{c})\Theta,
\end{equation}

\begin{equation} \label{eq10}
    \Delta N^{\mu}_{\hspace{1ex};\mu}=0,
\end{equation}

\noindent where $\Pi$ is the bulk viscous pressure, $p_{c}$ is the creation pressure, and $\Theta\equiv3H(a)$ is the expansion rate of the fluid. The creation pressure $p_{c}$ is associated with the adiabatic matter creation processes \cite{gunzig, prigogine1988, prigogine1989}. Considering that, in cosmology, we can relate the creation or destruction of particles to the bulk viscosity \cite{calvao, lima1992, zimdahl1993, zimdahl1996.2, lima1996}, the $p_{c}$-term becomes, to some extent, redundant. Thus, without loss of generality, 

\begin{equation} \label{eq11}
    U_{\mu}\Delta T^{\mu\nu}_{\hspace{1.5ex};\nu}=\Pi\Theta.
\end{equation}

Using Eq. (\ref{eq1}), the null divergence of $T^{\mu\nu}$ projected in the 4-velocity direction leads to the continuity equation:

\begin{equation} \label{eq12}
    U_{\mu}T^{\mu\nu}_{\hspace{1.5ex};\nu}=\dot{\rho}_{x}+3H(a)(\rho_{x}+p_{eq})=-3H(a)\Pi,
\end{equation}

\noindent where $p_{eq}\equiv w_{0}\rho_{x}$ is the equilibrium pressure and $\Pi=w_{a}f(a)\rho_{x}$ is the bulk viscous pressure mimicked by the dark energy fluid \cite{silva2012, silva2013}. According to Eq. (\ref{eq2}), the particle balance equation becomes \cite{costanetto}:

\begin{equation} \label{eq13}
    N^{\mu}_{\hspace{1ex};\mu}=\dot{n}+3H(a)n=n\Gamma.
\end{equation}

\noindent Consequently, the new entropy source of the component is \cite{costanetto}:

\begin{equation} \label{eq14}
    S^{\mu}_{\hspace{1ex};\mu}=n\dot{\sigma}+n\sigma\Gamma.
\end{equation}

\noindent Setting $S^{\mu}_{\hspace{1ex};\mu}\geq0$ in Eq. (\ref{eq14}), we get $\Gamma\geq-\dot{\sigma}/\sigma$. If $\dot{\sigma}=0$, as in the cosmological fluid \cite{gunzig, prigogine1988, prigogine1989}, $\Gamma\geq0$ and particles can only be created --- in this case, it is common to use $p_{c}$ instead of $\Pi$ in Eq. (\ref{eq11}). In the non-adiabatic scenario, we also get particle creation if $\dot{\sigma}<0$. Particle destruction, $\Gamma<0$, is possible as long as $\dot{\sigma}>0$. Since there is no special reason to take $\dot{\sigma}=0$ for dark energy, particles can be not only created but also destroyed, and this does not violate the second law of thermodynamics or any other. In fact, considering that in models with $\Gamma=0$ the source of entropy is $S^{\mu}_{\hspace{1ex};\mu}=n\dot{\sigma}$, assuming $\dot{\sigma}=0$ is a direct contradiction to the conclusion that the fluid leaves the adiabatic limit and begins to suffer from bulk viscosity when its barotropic EoS parameter varies with time \cite{silva2012, silva2013}. Therefore, \textit{a priori}, both processes of particle creation or destruction are possible.

Another relevant aspect concerns the effect of $\Gamma$ on the Hubble parameter and the possibility of constraining it observationally. The processes of matter creation in cosmology have been the subject of long-standing debate. In the case where the cosmological fluid satisfies $\dot{\sigma}=0$, only particle creation processes are allowed  \cite{gunzig, prigogine1988, prigogine1989}.  Under this condition, it is possible to derive a relation between the bulk viscous pressure and $\Gamma$ such that $\Gamma$ enters the continuity equation and, consequently, contributes explicitly to the Hubble parameter:

\begin{equation} \label{eq15}
    \Gamma=-\frac{3H(a)\Pi}{\rho_{x}+p_{eq}}.
\end{equation}

\noindent We emphasize, however, that Eq. (\ref{eq15}) is only valid in the context of the adiabatic creation of matter discussed in \cite{gunzig, prigogine1988, prigogine1989}. In the more general scenario we are interested in, $\dot{\sigma}\neq0$, there is no well-defined expression in theory for $\Gamma$ or relationships between $\Gamma$ and $\Pi$. In such cases, one must rely on phenomenological prescriptions. For example, in \cite{calvao} it is used

\begin{equation} \label{eq16}
    \Gamma=-\frac{1}{\beta}3H(a)\Pi,
\end{equation}

\noindent where $\beta$ is a constant. Although the above expression is an \textit{ansatz}, it takes into account, as pointed by the authors, particle creation processes in which the generated entities
are in thermal equilibrium with the energy-material content already present in the cosmological fluid, so that these processes become the sole cause of entropy production in the cosmos. Therefore, $\beta$ is defined as positive, so that $\Pi$ is negative. In our work, we sought, based on \cite{lima1996}, to adopt a more general functional form for $\Gamma$ that would not depend on scenarios such as the one described in \cite{gunzig, prigogine1988, prigogine1989} or the one assumed in \cite{calvao}. Thus, we use the expression given in Eq. (\ref{eq3}),

\begin{equation*} \label{eq17}
    \Gamma=3\alpha H(a).
\end{equation*}

\noindent This relation does not link $\Pi$ and $\Gamma$ and, therefore, does not allow the rate of particle creation or destruction to affect  the Hubble parameter. For this to be possible, it would be necessary to have a functional form for $\Gamma$ that relates this term to the energy-momentum tensor of dark energy, i.e., a functional form that connects it to the bulk viscous pressure without considering adiabatic scenarios. However, this is not the objective of this work.

Applying the functional form for the $\Gamma$-term given in Eq. (\ref{eq3}), the solutions to Eqs. (\ref{eq12}, \ref{eq13}) are, respectively \cite{silva2012, silva2013, costanetto}:

\begin{equation} \label{eq18}
    \rho_{x}=\rho_{x,0}a^{-3(1+w_{0})}\exp\left(-3w_{a}\int_{1}^{a} f(\tilde{a})\frac{d\tilde{a}}{\tilde{a}}\right)
\end{equation}

\noindent and

\begin{equation} \label{eq19}
    n=n_{0}a^{-3(1-\alpha)}.
\end{equation}

\noindent In \cite{costanetto}, we verified that the temperature evolution law presented in \cite{silva2012, silva2013} continues to hold. Then

\begin{equation} \label{eq20}
    \frac{\dot{T}}{T}=-3H(a)\left[\left(\frac{\partial p_{0}}{\partial \rho_{x}}\right)_{n}+\left(\frac{\partial \Pi}{\partial \rho_{x}}\right)_{n}\right]
\end{equation}

\noindent and, therefore,

\begin{equation} \label{eq21}
    T=T_{0}a^{-3w_{0}}\exp\left(-3w_{a}\int_{1}^{a} f(\tilde{a})\frac{d\tilde{a}}{\tilde{a}}\right).
\end{equation}

\noindent Assuming that the relation $\mu/T=\mu_{0}/T_{0}$ \cite{lima2008, pereira} remains valid in the non-equilibrium context, the Euler relation for thermodynamics leads us to the specific entropy of the fluid \cite{costanetto}:

\begin{equation} \label{eq22}
    \sigma=\frac{\rho_{x,0}}{n_{0}T_{0}}(1+w_{0}+w_{a}f(a))\frac{1}{a^{3\alpha}}-\frac{\mu_{0}}{T_{0}}.
\end{equation}

\noindent Thus,

\begin{equation} \label{eq23}
    \dot{\sigma}=\frac{\rho_{x,0}}{n_{0}T_{0}}\left[w_{a}af'(a)-3\alpha(1+w_{0}+w_{a}f(a))\right]\frac{H(a)}{a^{3\alpha}}.
\end{equation}

\noindent Note that $\dot{\sigma}$ can mathematically be either positive or negative, as we expected. The entropy density is:

\begin{equation} \label{eq24}
    s=\left[s_{0}+\rho_{x,0}(w(a)-w_{0})-\mu_{0}n_{0}(a^{3\alpha}-1)\right]\frac{1}{T_{0}a^{3}}.
\end{equation}

\noindent With these quantities determined, we can obtain the entropy and entropy source of dark energy in out-of-equilibrium models.

Since $S=(n\sigma)V$ is the total entropy, where $V=V_{0}a^{3}$ is the comoving volume \cite{weinberg1972, weinberg2008}, we get \cite{costanetto}:

\begin{equation} \label{eq25}
    S=\left[\rho_{x,0}(1+w_{0}+w_{a}f(a))-\mu_{0}n_{0}a^{3\alpha}\right]\frac{V_{0}}{T_{0}}.
\end{equation}

\noindent Substituting Eqs. (\ref{eq19}, \ref{eq22}, \ref{eq23}) into Eq. (\ref{eq14}), the entropy source of the fluid is obtained:

\begin{equation} \label{eq26}
    S^{\mu}_{\hspace{1ex};\mu}=\left[w_{a}f'(a)\rho_{x,0}-3\alpha\mu_{0}n_{0}a^{3\alpha-1}\right]\frac{H(a)}{T_{0}a^{2}}.
\end{equation}

\noindent Before getting to the thermodynamic constraint equations for $\mu_{0}n_{0}$, it is interesting to note that $S\geq0$ leads us to $w(a)_{min}=-1+(\mu_{0}n_{0}/\rho_{x,0})a^{3\alpha}$. As we can see, if $\mu<0$, then $w(a)_{min}<-1$ and dark energy is in the phantom regime; if $\mu=0$, then $w_{min}=-1$ and the cosmological constant scenario is recovered; if $\mu>0$, then $w(a)_{min}>-1$.

Imposing the positivity of entropy on Eq. (\ref{eq25}), we arrive at an upper limit for the product of the chemical potential and the particle density:

\begin{equation} \label{eq27}
    \mu_{0}n_{0}\leq(1+w_{0}+w_{a}f(a))\frac{\rho_{x,0}}{a^{3\alpha}}.
\end{equation}

\noindent From the second law of thermodynamics, 

\begin{equation} \label{eq28}
    \alpha\mu_{0}n_{0}\leq\frac{w_{a}f'(a)}{3a^{3\alpha-1}}\rho_{x,0}.
\end{equation}

\noindent We have two possibilities depending on the sign of $\alpha$: if it is positive, we will obtain another upper bound for $\mu_{0}n_{0}$ different from the one derived from $S\geq0$; if it is negative, Eq. (\ref{eq26}) will lead us to a lower limit for this product. In summary, we can write:

\begin{equation} \label{eq29}
    \mu_{0}n_{0}
    \begin{cases}
        \leq\dfrac{w_{a}f'(a)}{3\alpha a^{3\alpha-1}}\rho_{x,0}, \,\,\, \alpha>0,
        \\
        \\
        \geq\dfrac{w_{a}f'(a)}{3\alpha a^{3\alpha-1}}\rho_{x,0}, \,\,\, \alpha<0.
    \end{cases}
\end{equation}

\noindent As we can see, thermodynamics will not necessarily produce upper and lower bounds for the product $\mu_{0}n_{0}$. This will depend on the sign of the constant $\alpha$. In any case, Eqs. (\ref{eq27}, \ref{eq29}) are sufficient to impose observational constraints and restrict the possible values for this quantity.
\section{Observational data and statistical analyses} \label{sec3}

As demonstrated in Section \ref{sec2}, the product of the chemical potential and the number density of dark energy is constrained by thermodynamic considerations. The positivity of the entropy defines an upper bound ($l_{\mathrm{sup}}^{\mathrm{PE}}$) for $\mu_0 n_0$, while the second law of thermodynamics imposes an additional constraint: an upper bound ($l_{\mathrm{sup}}^{\mathrm{SL}}$) for positive values of $\alpha$ and a lower bound ($l_{\mathrm{inf}}^{\mathrm{SL}}$) for negative values of $\alpha$.  These thermodynamic limits depend on the redshift, the parameter $\alpha$ and cosmological parameters ($\Omega_m, H_0, w_0, w_a$), which are statistically constrained by cosmological probes.  In order to constrain $\mu_0 n_0$, we perform statistical analyses under the assumption of a spatially flat universe in which the dark energy component is modeled as a fluid characterized by the CPL and BA EoS. The following cosmological data are considered:

\begin{figure*}[t!]
    {\includegraphics[width=0.46\textwidth]{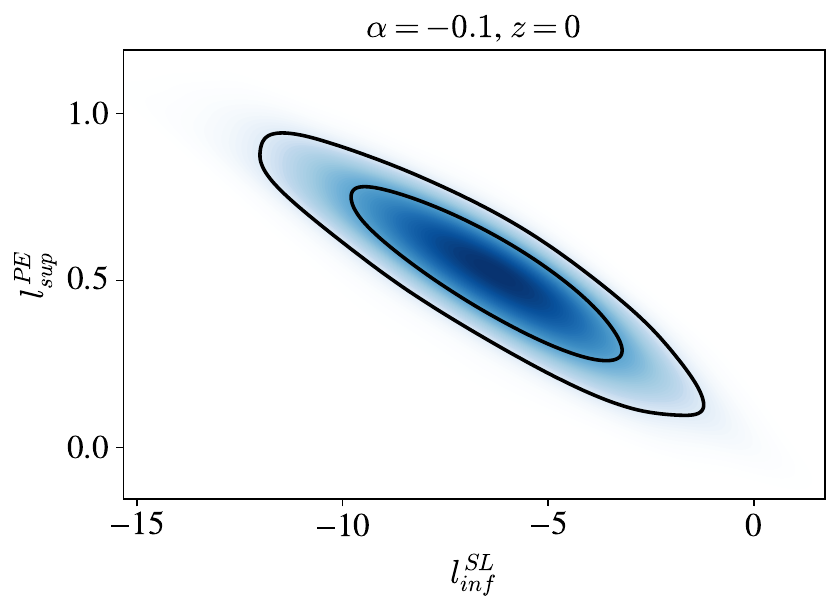}}
    {\includegraphics[width=0.46\textwidth]{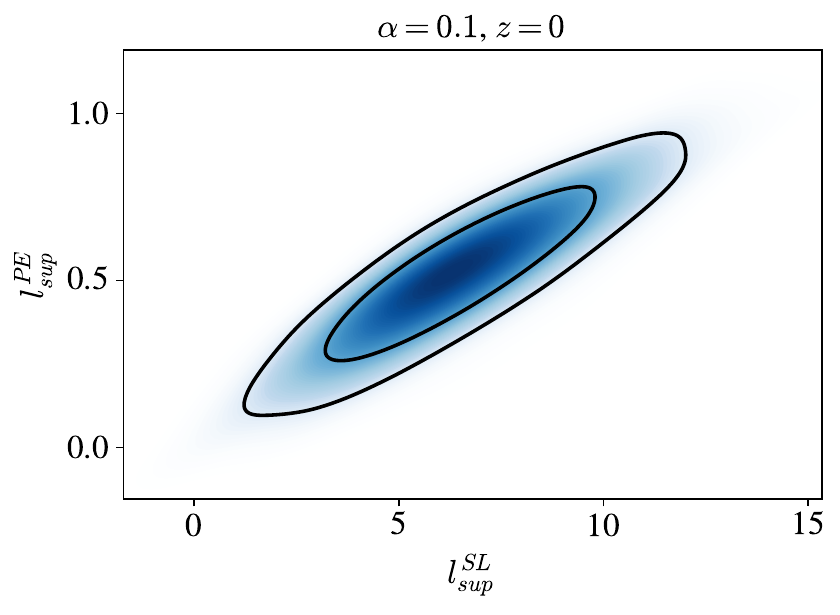}}
    {\includegraphics[width=0.46\textwidth]{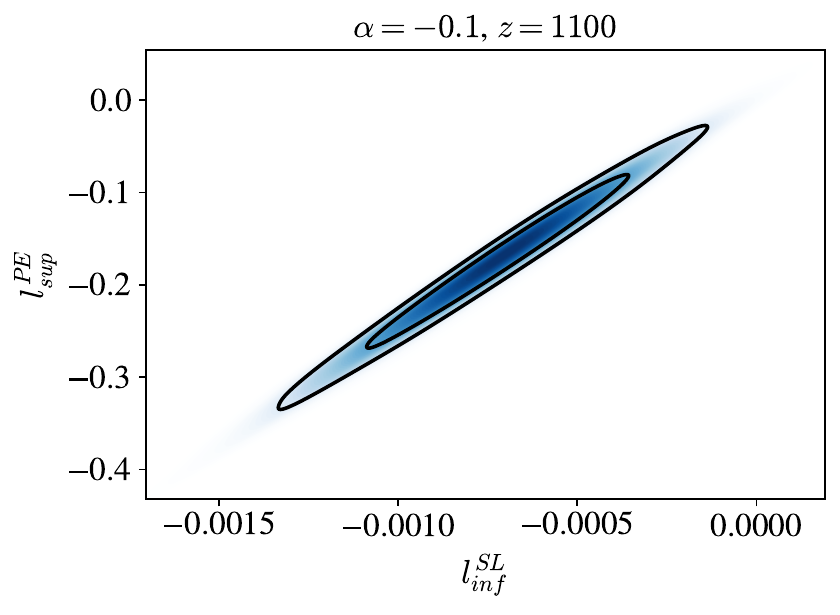}}
    {\includegraphics[width=0.46\textwidth]{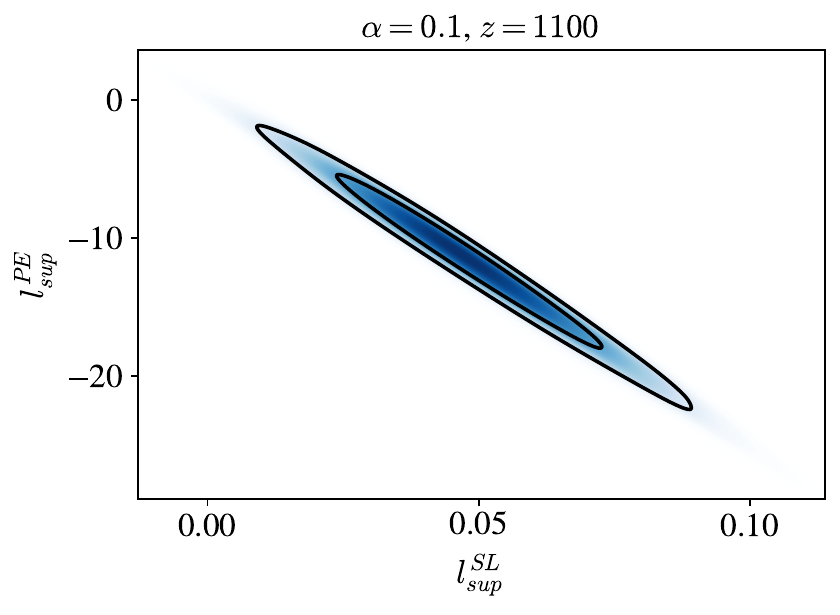}}
    \caption{1$\sigma$ and 2$\sigma$ confidence levels of the thermodynamic limits for two values of the parameter $\alpha$ and two redshifts in the CPL model.}
    \label{fig:constraints}
\end{figure*}

\begin{itemize}
    \item \textbf{SN Ia:} The Pantheon+ compilation \cite{pantheon2022.1, pantheon2022.2} is employed, comprising $1701$ light curves from $1550$ spectroscopically confirmed supernovae, covering a redshift range from $0.01$ up to $2.34$. The apparent magnitude $m$ serves as the relevant cosmological observable, being related to cosmic dynamics through the luminosity distance. The high quality and large volume of this dataset enable stringent constraints on the two-parameter dark energy models considered.
    
    \item \textbf{BAO:}  Using approximately 14 million galaxy positions and redshifts, the DESI Collaboration has measured BAO signals with unprecedented precision \cite{desi2}. Based on these data, five tracers\footnote{Bright Galaxies (see \cite{desi2} for details of the sample), Luminous Red Galaxies, Emission Line Galaxies, Quasars, and the Ly-$\alpha$ Forest.} were employed to determine six independent measurements of the anisotropic BAO signal, yielding corresponding estimates of $D_M/r_d$ and $D_H/r_d$ over an effective redshift range of $0.510$ to $2.330$, as well as one isotropic measurement providing $D_V/r_d$ at $z=0.295$.
    
    \item \textbf{CMB}: The temperature and polarization fluctuations in the CMB constitute one of the most powerful cosmological probes. Measurements from the Planck satellite \cite{planck2018.1, planck2018.2}, in combination with other observational datasets, enable precise constraints on the energy density components of the universe and on the parameters of a two-parameter dark energy EoS. We employ the temperature (TT), temperature–polarization (TE), and polarization (EE) power spectra, adopting the \textit{Commander} likelihood for the temperature spectrum and the \textit{SimAll} likelihood for the polarization spectrum at low multipoles ($l>30$), and the most recent \textit{CamSpec} PR4 likelihood for the TT, TE, and EE spectra at high multipoles ($l>30$) \cite{Efstathiou:2019mdh, Rosenberg:2022sdy}. In addition, we include the CMB lensing reconstruction power spectrum from the PR4 data \cite{carron}.
\end{itemize}

\begin{table}[t!]
    \centering
    \begin{tabular}{c c}
         \hline\\
          Parameter & Prior \\
         \hline 
         \rule{0pt}{3ex} 
    
        $\omega_{b,0}$ &$\mathcal{U}[0.005,0.1]$\\
         $\omega_{c,0}$ & $\mathcal{U}[0.001,0.99]$\\
         $w_{0}$ & $\mathcal{U}[-3,1]$\\
         $w_{a}$ & $\mathcal{U}[-3,2]$\\
           $\ln 10^{10}A_s$ & $\mathcal{U}[1.61,3.91]$ \\
          $n_s$ & $\mathcal{U}[0.8,1.2]$\\
         $100\theta_s$& $\mathcal{U}[0.5,10]$\\

         $\tau$ & $\mathcal{U}[0.01,0.8]$\\
         \hline 
    
    \end{tabular}
    \caption{Priors used for the MCMC analysis considering Pantheon+, DESI DR2, and Planck PR4 data. We assume uniform priors which are noted as $\mathcal{U}$[Min; Max].}
    \label{tab:priors}
\end{table}

To constrain the parameters of the aforementioned models, we use the \textit{COBAYA} code \cite{COBAYA} to perform Bayesian analyses through Markov chain Monte Carlo (MCMC) from the Metropolis-Hasting sampler. We use the \textit{CLASS} code \cite{CLASS} and a modification of it to compute the theoretical cosmological  equations. Following the performed analysis in \cite{desi2}, the assumed priors are presented in Table \ref{tab:priors}, where $\omega_{b,0}$ is the physical baryon density, $\omega_{c,0}$ is the physical dark matter density, $w_0$ and $w_a$ are the usual parameters of the dark energy EoS, $\theta_s$ is the angular size of the sound horizon at recombination, $A_s$ is the scalar amplitude, $n_s$ is the spectral index and $\tau$ is the optical depth to reionization.

Table ~\ref{tab:posteriors} summarizes the results of the statistical analyses of cosmological parameters related to the thermodynamic constraints for the CPL and BA parameterizations. This analysis was performed using the \textit{GetDist} package \cite{GetDist}.

\begin{table}[h!]
    \centering
    \begin{tabular}{c c c}
         \hline\\
          Parameter & CPL& BA \\
         \hline 
         \rule{0pt}{3ex} 
    
        $\Omega_{m,0}$ &$0.3109\pm 0.0058$ & $0.3104\pm 0.0058$\\
         $H_{0}$ & $67.51^{+0.58}_{-0.61}$& $67.59\pm 0.61$\\
         $w_{0}$ & $-0.850^{+0.053}_{-0.052}$&$-0.873^{+0.059}_{-0.046}$ \\
         $w_{a}$ &$-0.56^{+0.20}_{-0.21} $& $-0.28^{+0.099}_{-0.10}$\\
         \hline 
    
    \end{tabular}
    \caption{Marginalized results of the cosmological parameters that define the thermodynamic limits in Eqs. (\ref{eq27}, \ref{eq29}).}
    \label{tab:posteriors}
\end{table}

Since the thermodynamic limits depend on the cosmological parameters, they can be regarded as random variables whose distributions, $P(l^{\mathrm{SL}},l^{\mathrm{PE}})$, are derived from the posterior distributions of the cosmological parameters \{$\Omega_m, H_0, w_0, w_a$\}, and the $\alpha$ and the redshift values. In Figure \ref{fig:constraints}, we present the bidimensional distributions of these limits for four combinations of values of $\alpha$ and $z$ considering the CPL model. The results of the thermodynamic limits and the constraints on the product $\mu_0 n_0$ for the BA parameterization are presented in Appendix \ref{ap:BA}.
\section{Observational limits on the $\mu_{0}n_{0}$ product} \label{sec4}

\begin{figure}[t!]
    {\includegraphics[width=0.48\textwidth]{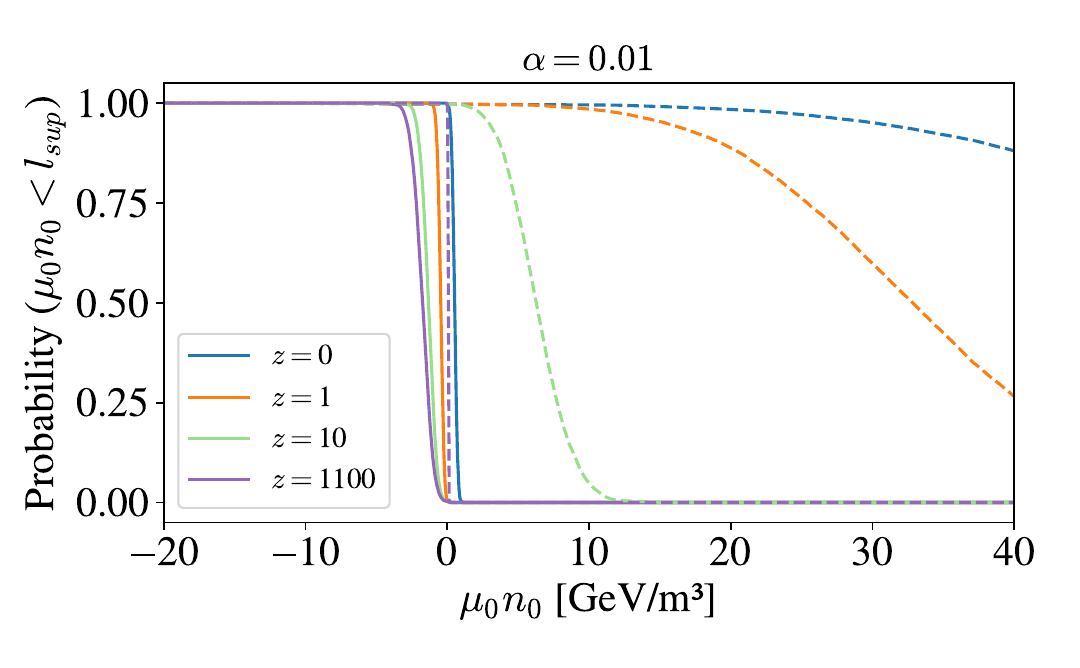}}
    {\includegraphics[width=0.48\textwidth]{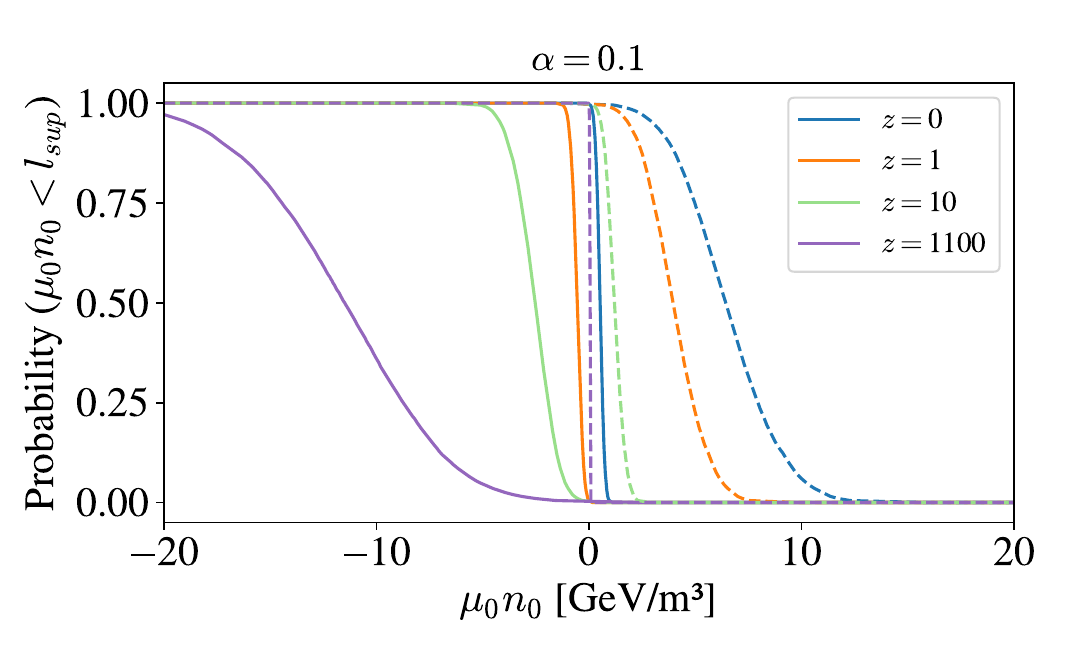}}
    \caption{Probability that a given value of $\mu_0 n_0$ satisfies the thermodynamic conditions for positive values of $\alpha$ in the CPL model. The solid lines correspond to the bound of the positivity of the entropy and the dashed ones to the second law of thermodynamics.}
    \label{fig:positive_limits}
\end{figure}

\begin{figure*}[t!]
    {\includegraphics[width=0.49\textwidth]{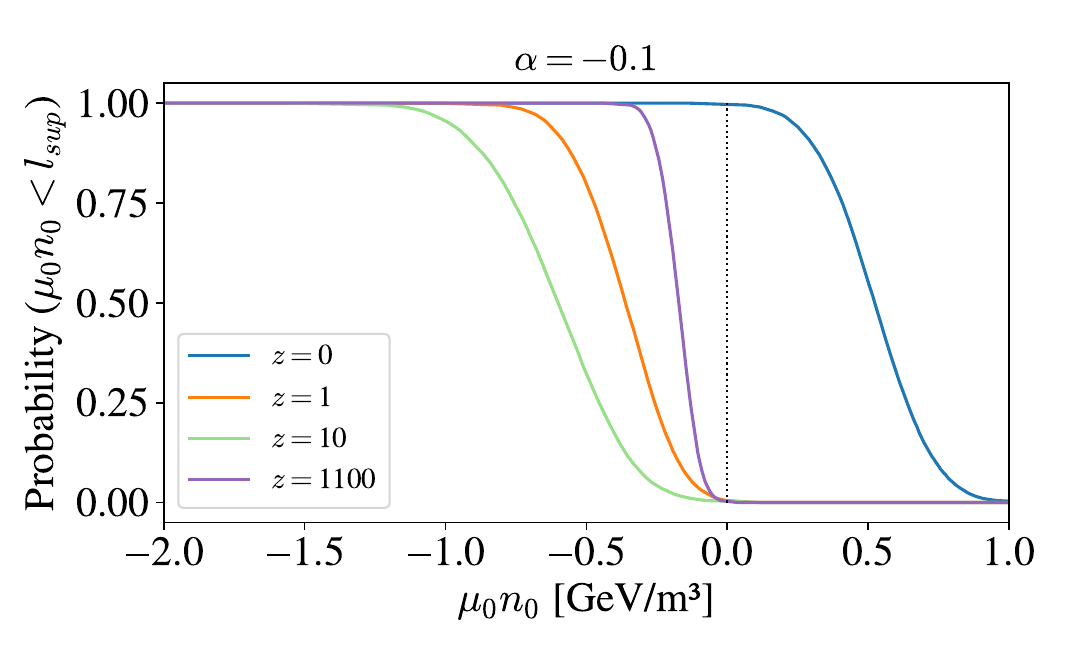}}
    {\includegraphics[width=0.49\textwidth]{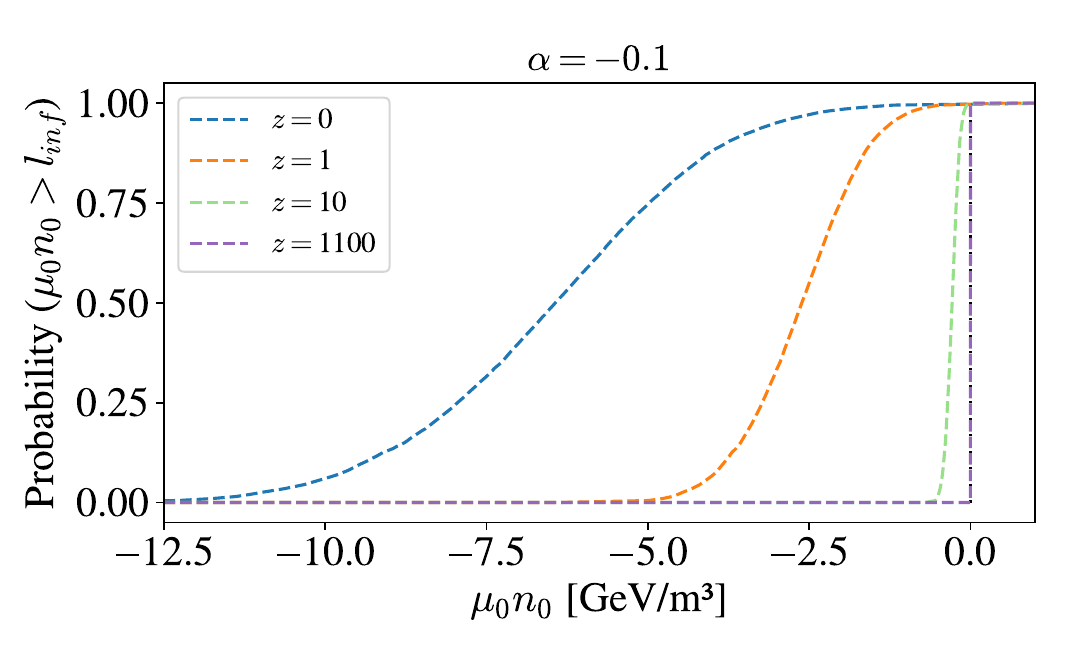}}
    {\includegraphics[width=0.49\textwidth]{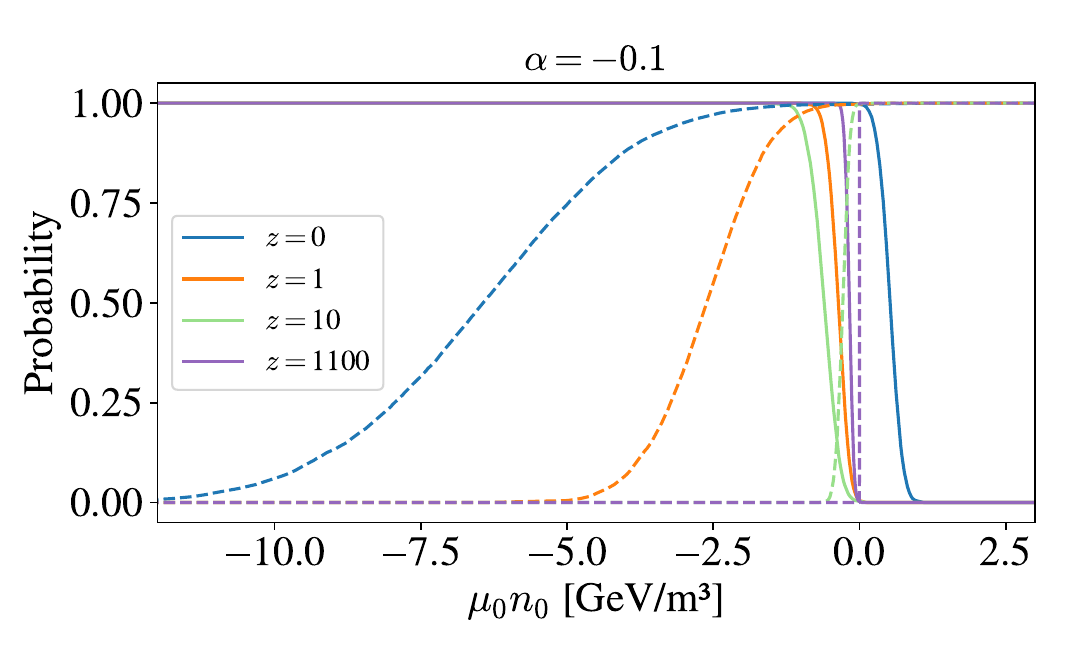}}
    {\includegraphics[width=0.49\textwidth]{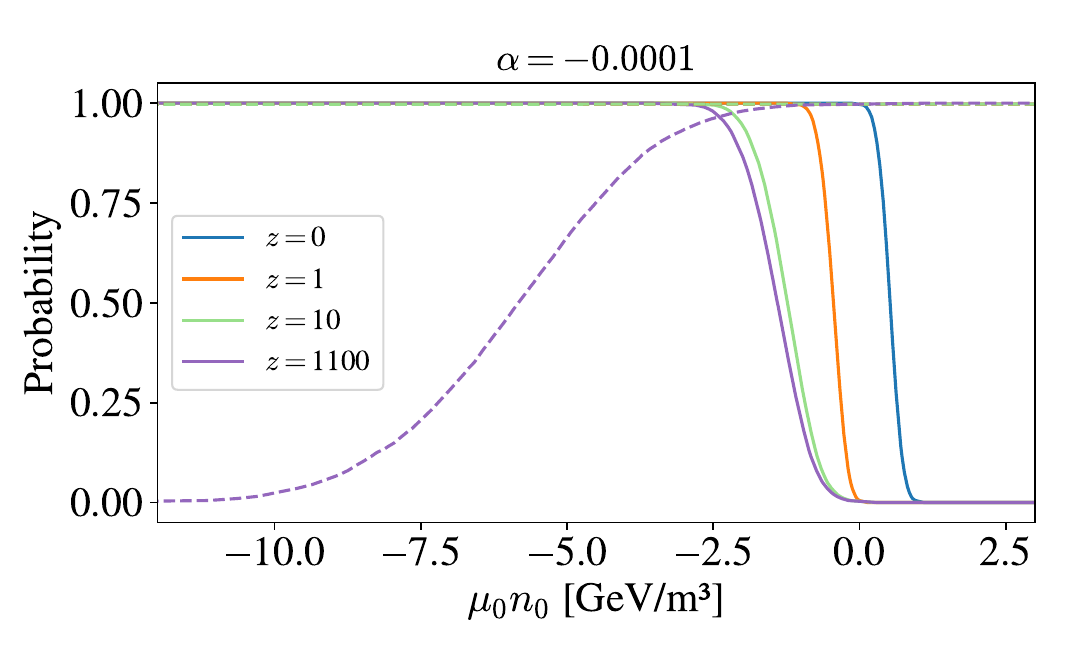}}
    \caption{Probability that a given value of $\mu_0 n_0$ satisfies the thermodynamic conditions for negative values of $\alpha$ in the CPL model. The solid lines correspond to the bound of the positivity of the entropy and the dashed ones to the second law of thermodynamics. The dotted line represents the zero limit.}
\label{fig:negative_limits}
\end{figure*}

Since the product $\mu_0 n_0$ and the cosmological parameters are related through an inequality rather than a functional dependence, it is convenient to represent the cumulative probability that a given value of $\mu_0 n_0$ satisfies the corresponding upper or lower thermodynamic constraint. Mathematically, this can be written as

 \begin{equation} \label{eq30}
     P(\mu_0 n_0 < l_{sup})=\int_{\mu_0 n_0}^\infty P(l_{sup})dl_{sup}
 \end{equation}

\noindent or 

\begin{equation} \label{eq31}
     P(\mu_0 n_0 > l_{inf})=\int^{\mu_0 n_0}_{-\infty} P(l_{inf})dl_{inf},
 \end{equation}

\noindent where $P(l_{sup})$ and $P(l_{inf})$ denote the marginal probability distributions of the upper and lower thermodynamic bounds, respectively. These cumulative probabilities for positive and negative values of $\alpha$ assuming the CPL model are shown in Figures \ref{fig:positive_limits} and \ref{fig:negative_limits}, respectively.

\subsection{$\alpha>0$}

As shown in Figure \ref{fig:positive_limits}, the probability distribution of the set of cosmological parameters \{$\Omega_m, H_0, w_0, w_a$\} yields limits where the requirement of positive entropy (solid lines) is more restrictive than the constraint imposed by the second law of thermodynamics (dashed lines). This behavior is also evident in the right panels of Figure \ref{fig:constraints}, where the limits associated with the second law consistently lie at higher values in these distributions. Another important feature is that the most restrictive limits arise from the condition of Eq. (\ref{eq27}) and the first condition of Eq. (\ref{eq29}) in the highest redshift of the data ($z=1100$ corresponding to the CMB data). Moreover, an increasing in the value of $\alpha$ further tightens the allowed range of $\mu_0 n_0$. 

Since in the cases where $\alpha >0$ both thermodynamic bounds act solely as upper limits, they effectively do not constrain the $\mu_{0}n_{0}$ product of dark energy. Instead, they impose bounds on their possible values. Knowing that the number density is always positive, from Figure \ref{fig:positive_limits} it follows directly that the present time value of the fluid chemical potential must be negative, $\mu_{0}<0$. Since $\mu/T$ is assumed to be a constant quantity and bearing in mind that $T_{0}/T$ is also always positive, $\mu_{0}<0$ implies $\mu<0$, suggesting a preference for the phantom regime when $\alpha>0$.

\subsection{$\alpha<0$}

Scenarios with negative values of $\alpha$ are particularly interesting due to the existence of both upper and lower thermodynamic bounds. In these cases, the second law of thermodynamics implies the second condition in Eq. (\ref{eq29}), thereby constraining the possible values of the $\mu_{0}n_{0}$ product both from below and above. Figure \ref{fig:negative_limits} shows the cumulative probability of a given value of $\mu_0 n_0$ to satisfy the thermodynamic constraints. In the first row of panels, the limits for $\alpha = -0.1$ are displayed independently. When these two conditions are combined (lower-left panel), it becomes evident that no value of $\mu_0 n_0$ satisfy them simultaneously. It is worth mentioning that all limits must be satisfied throughout the entire redshift range of the data; therefore, the most restrictive bounds are adopted, that is, the highest value of the lower limits and the lowest value of the upper limits.

The scenario changes as the parameter $\alpha$ approaches zero from the left. In the low-right panel of Figure \ref{fig:negative_limits}, a compatibility region appears for $\alpha=-0.0001$, where the effective bounds are determined by the restriction at $z=1100$. We analyze the behavior of the second law of thermodynamics and the positivity of entropy limits for various negative values of $\alpha$ and find a minimum value of approximately $\alpha=-0.0002$ that allows compatibility between the two thermodynamic bounds. This result establishes a lower limit on $\alpha$ and consequently constrains the possible particle destruction rate in dark energy. For the cases where the values of $\alpha$ allow compatibility between both thermodynamic bounds, the product of the chemical potential and the number density can be treated as a random variable, and its probability distribution can be computed as follows.

From the definition of the conditional probability of $\mu_0 n_0$ given the lower and upper limits, $P(\mu_0 n_0|l_{\mathrm{inf}}^{\mathrm{SL}},l_{\mathrm{sup}}^{\mathrm{PE}})$, which relates the joint probability  $P(\mu_0 n_0,l_{\mathrm{inf}}^{\mathrm{SL}},l_{\mathrm{sup}}^{\mathrm{PE}})$ to the posterior probability of the thermodynamic limits $P(l_{\mathrm{inf}}^{\mathrm{SL}},l_{\mathrm{sup}}^{\mathrm{PE}})$, we have

\begin{equation} \label{eq32}
     P(\mu_0 n_0,l_{\mathrm{inf}},l_{\mathrm{sup}})=P(\mu_0 n_0|l_{\mathrm{inf}},l_{\mathrm{sup}})P(l_{\mathrm{inf}},l_{\mathrm{sup}}).
\end{equation}

\noindent The  probability distribution of $\mu_0 n_0$, $P(\mu_0 n_0)$, can then be obtained marginalizing over $l_{inf}$ and $l_{sup}$,  
 
\begin{align} \label{eq33}
    P(\mu_0 n_0) &= \int   P(\mu_0 n_0,l_{\mathrm{inf}},l_{\mathrm{sup}})\, d l_{\mathrm{inf}}\, d l_{\mathrm{sup}} \nonumber \\
    &= \int P(\mu_0 n_0|l_{\mathrm{inf}},l_{\mathrm{sup}})\, P(l_{\mathrm{inf}},l_{\mathrm{sup}})\, d l_{\mathrm{inf}}\, d l_{\mathrm{sup}}.
\end{align}

\begin{figure}[t!]
    {\includegraphics[width=0.46\textwidth]{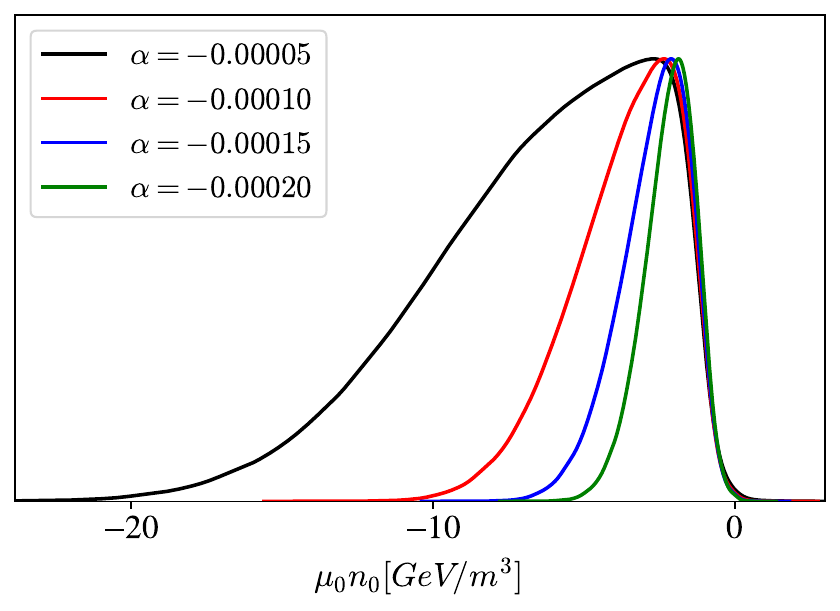}} 
    \caption{Probability distributions of $\mu_0 n_0$ product for negative values of $\alpha$.}
    \label{fig:marginal_pdf}
\end{figure}

\noindent For this purpose, it is necessary to define the conditional probability $P(\mu_0 n_0|l_{\mathrm{inf}}^{\mathrm{SL}},l_{\mathrm{sup}}^{\mathrm{PE}})$. Due to the uncertain nature of dark energy, we assume a flat distribution of $\mu_0 n_0$ conditioned on any pair of limits \{$l_{inf}-l_{sup}$\}, considering only those pairs that satisfy $l_{inf}<l_{sup}$. To estimate the marginalized probability distribution $P(\mu_0 n_0)$, we use the chain provided by the MCMC process in the statistical analysis as a sample of the probability distribution $P(l_{\mathrm{inf}},l_{\mathrm{sup}})$  (as shown in Figure \ref{fig:constraints}). We then perform a Monte Carlo sampling of $\mu_0 n_0$ by drawing from uniform conditional distributions $P(\mu_0 n_0|l_{\mathrm{inf}}^{\mathrm{SL}},l_{\mathrm{sup}}^{\mathrm{PE}})$ for each pair of thermodynamic limits \{$l_{inf}-l_{sup}$\}. The resulting  marginalized probability distributions of $\mu_0 n_0$ for various negative values of $\alpha$, including its lowest allowed value, are presented in Figure \ref{fig:marginal_pdf}. A summary of these distributions, along with their median values and $1\sigma$ confidence levels, is provided in Table \ref{tab:results}.

\begin{table}[h!]
    \centering
    \begin{tabular}{c  c}
         \hline
          $\alpha$ & $\mu_0 n_0$  \\
           &  [GeV/m$^{3}$] \\
         \hline 
         \rule{0pt}{3ex} 
         $-0.00005$ & $ -6.6^{+5.2}_{-2.0}$\\
         $-0.00010$ & $-3.7^{+2.3}_{-1.1}$\\
         $-0.00015$ & $-2.7^{+1.4}_{-0.8}$\\
         $-0.00020$ & $-2.2^{+1.0}_{-0.7}$\\
         \hline 
    
    \end{tabular}
    \caption{Constraints on the $\mu_0 n_0$ product considering Pantheon+, DESI DR2, and Planck PR4 data for different values of the parameter $\alpha$.}
    \label{tab:results}
\end{table}

It follows that $\mu_{0}n_{0}(\alpha=-0.0002)=-2.2^{+1.0}_{-0.7}\,\,GeV/m^{3}$ is the value of this product for the limiting case of compatibility of the thermodynamic constraints. This quantity becomes more negative as $\alpha$ approaches zero from the left. Therefore, based on arguments similar to those presented in the previous case, the conclusion that the chemical potential must be negative is immediate, indicating a preference for a phantom dark energy when particles are being destroyed in the fluid.
\section{Conclusions} \label{concl}

The chemical potential plays a unique role in dark energy thermodynamics. Whether the component is in equilibrium or not, $\mu<0$ indicates that the fluid is in the phantom regime; $\mu=0$ recovers the cosmological constant; $\mu>0$ returns neither of the two cases \cite{lima2008, pereira}. By imposing observational constraints on the product of the chemical potential and the number density based on thermodynamic constraints derived from the positivity of entropy and the second law of thermodynamics, the research developed in this work showed that, regardless of the sign of $\alpha$ and considering that $n_{0}$ must always be positive, the present time value of the chemical potential of dark energy must be negative, $\mu_{0}<0$. Given that $\mu/T$ is considered constant and that $T>0$, this leads to the conclusion that the chemical potential must be negative not only in the present, but throughout cosmic history. Therefore, whether particles are being created or destroyed in the component and independently of the cosmic era, we observe $\mu<0$. This is a strong conclusion, and one that we intend to explore further in a future paper. For now, the results of this current work show a favoring of dark energy models behaving as a phantom fluid, a result compatible with the dark energy behavior at high-$z$ found by the DESI collaboration \cite{desi1, desi2}.

There is, apparently, as seen, no preference regarding the presence of a source or sink of particles in the fluid: in both cases, the chemical potential of dark energy is negative. However, from a theoretical point of view, it is possible to lean towards processes in which particles are destroyed --- as we have already argued in \cite{costanetto}. For example, only for $\alpha<0$ can we impose both an upper and a lower limit on the $\mu_{0}n_{0}$ product, so that, in this scenario, it is possible to establish a range of values for this product. Still within the context of particle destruction, there is a strong indication that the absolute values of $\alpha$ must be very small due to the compatibility of the thermodynamic constraints, according to what is seen in \cite{lima1999.1, lima1999.2}. From this, we establish $\alpha=-0.0002$ as an approximate limit value from which the thermodynamic constraints are compatible and $\mu_{0}n_{0}(\alpha=-0.0002)=-2.2^{+1.0}_{-0.7}\,\,GeV/m^{3}$ as the corresponding value for the product of dark energy chemical potential and number density.
\begin{acknowledgments}
     We dedicate this work to the memory of Professor Raimundo Silva, whose contributions to the field of dark energy thermodynamics deeply influenced this study and many others in the field. His scientific vision, intellectual rigor, and generosity as a mentor continue to inspire our research and the wider community. JG acknowledges support from the Foundation for Research and Technological Innovation Support of the State of Sergipe (FAPITEC/SE) under grant number 794017/2013.
\end{acknowledgments}
\appendix
\section{BA parameterization results} \label{ap:BA}

\begin{figure*}[t!]
    {\includegraphics[width=0.46\textwidth]{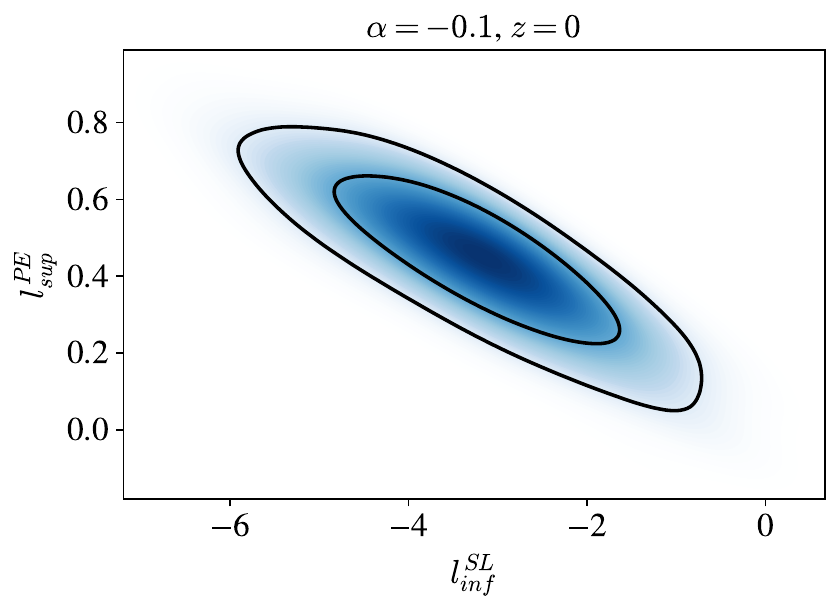}}
    {\includegraphics[width=0.46\textwidth]{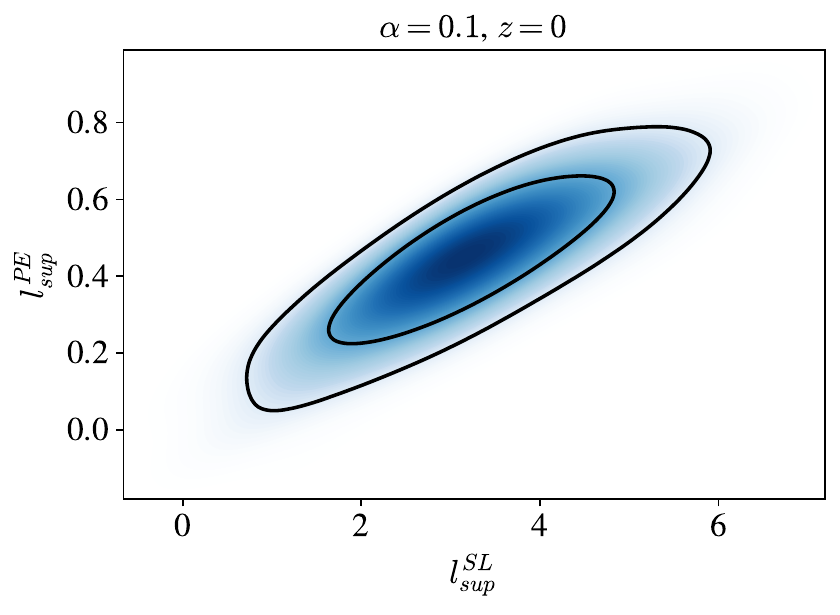}}
    {\includegraphics[width=0.46\textwidth]{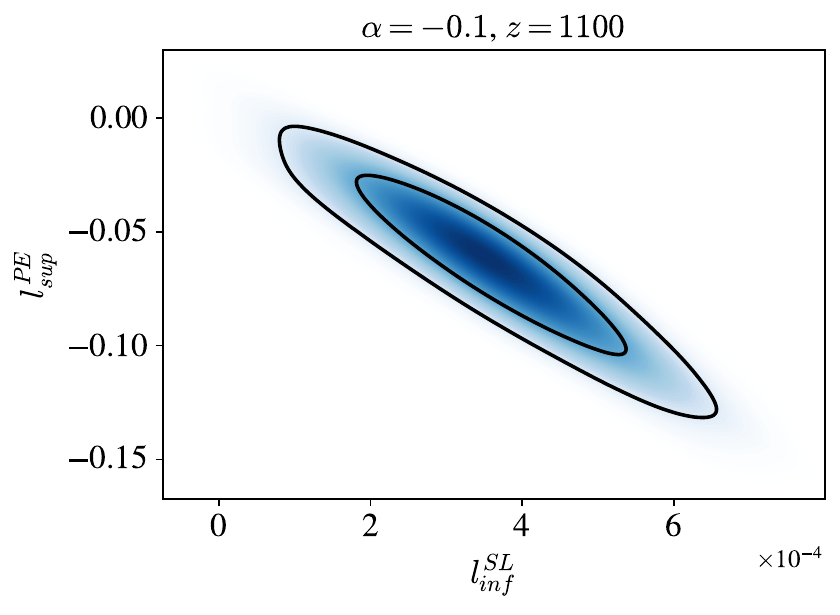}}
    {\includegraphics[width=0.46\textwidth]{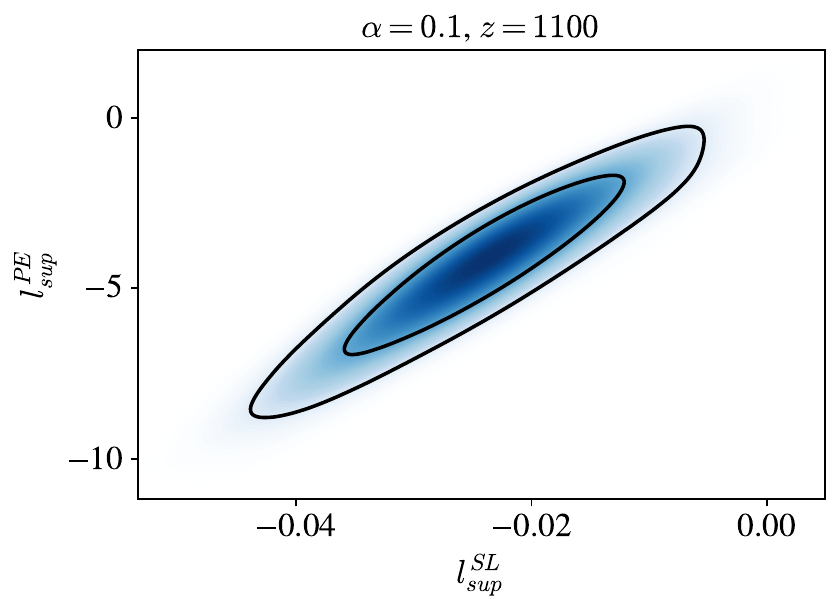}}
    \caption{1$\sigma$ and 2$\sigma$ confidence levels of the thermodynamic limits for two values of the parameter $\alpha$ and two redshifts in the BA model.}
    \label{fig:constraintsBA}
\end{figure*}

\begin{figure*}[t!]
    {\includegraphics[width=0.49\textwidth]{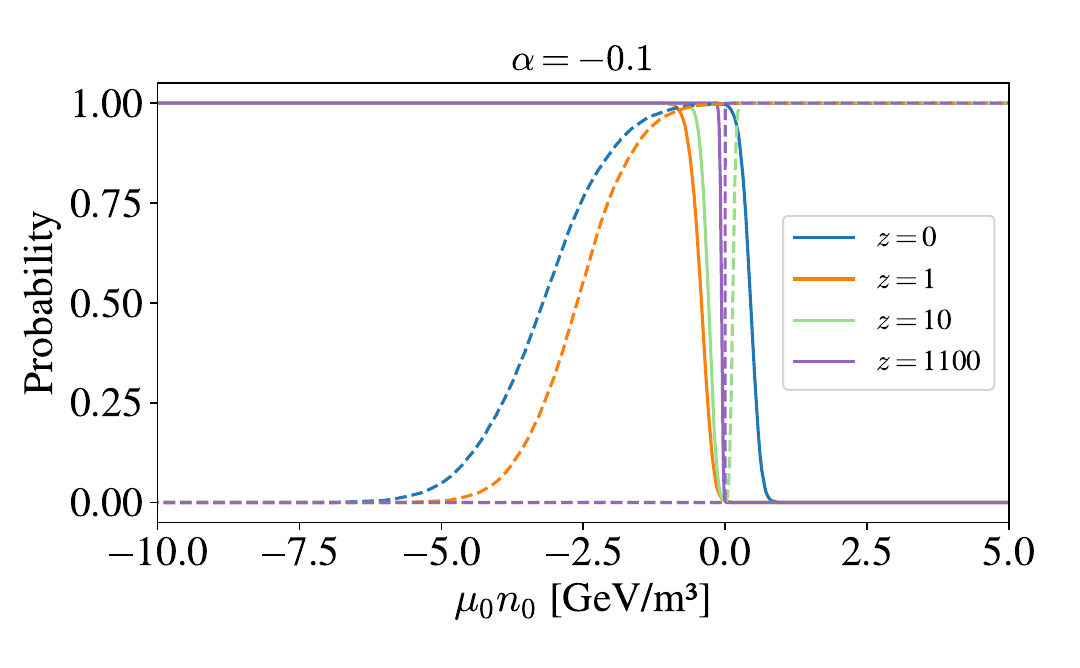}}
    {\includegraphics[width=0.49\textwidth]{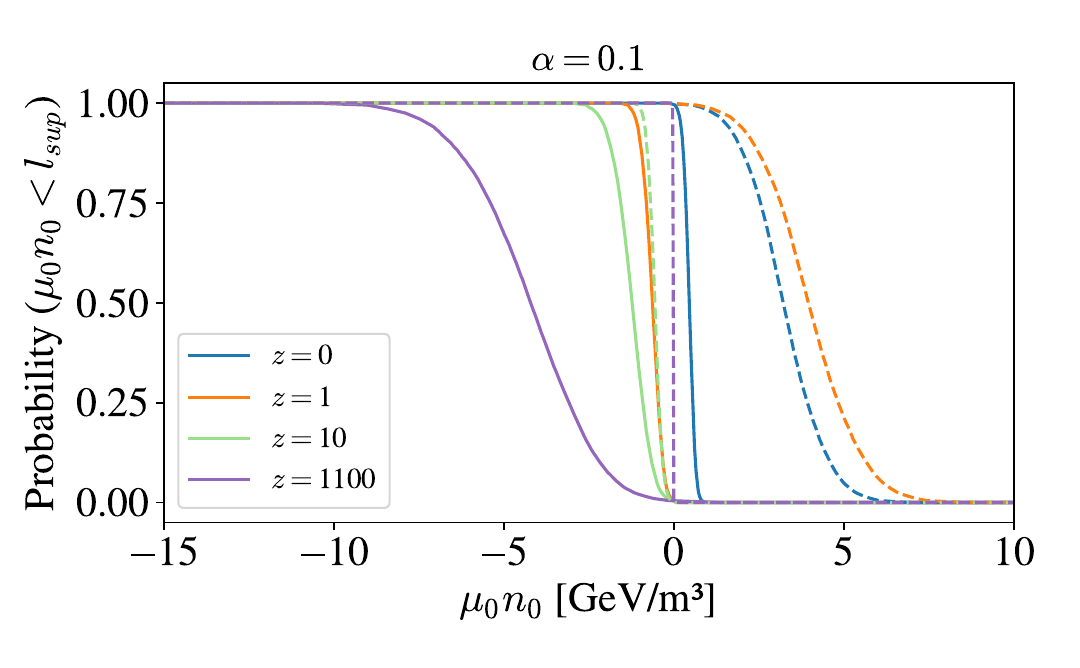}}
    \caption{Probability that a given value of $\mu_0 n_0$ satisfies the thermodynamic conditions for two values of $\alpha$ in the BA model. The solid lines correspond to the bound of the positivity of the entropy and the dashed ones to the second law of thermodynamics.}
    \label{fig:BA_limits}
\end{figure*}

The difference between the BA and CPL parameterizations leads to significant differences in the corresponding thermodynamic limits. As shown in Figure \ref{fig:constraintsBA}, the correlations between the limits exhibit an opposite behavior at $z=1100$ in contrast to the CPL results presented in Figure \ref{fig:constraints}. In Figure \ref{fig:BA_limits}, we present the cumulative probability that a given value of $\mu_0 n_0$ satisfies the thermodynamic bounds. Unlike the CPL parameterization, in the BA model no value of $\alpha$ allows simultaneous compatibility between the positivity of entropy and the second law of thermodynamics. This incompatibility prevents the calculation of the probability distribution of $\mu_0 n_0$.

\bibliographystyle{apsrev4-1}
\bibliography{refe}

@article{desi1,
    title={{DESI 2024 VI: Cosmological constraints from the measurements of baryon acoustic oscillations}},
    author={Adame, A G and others},
    journal={J. Cosmol. Astropart. Phys.},
    volume={2025},
    number={2},
    pages={071}, 
    year={2025},
    publisher={IOP Publishing},
    DOI={10.1088/1475-7516/2025/02/021}
}

@article{desi2,
    title={{DESI DR2 Results II: Measurements of Baryon Acoustic Oscillations and Cosmological Constraints}}, 
    author={Abdul-Karim, M and others},
    year={2025},
    eprint={2503.14738},
    archivePrefix={arXiv},
    primaryClass={astro-ph.CO},
    doi={10.48550/arXiv.2503.14738}
}

@article{Rosenberg:2022sdy,
    author = "Rosenberg, Erik and Gratton, Steven and Efstathiou, George",
    title = "{CMB power spectra and cosmological parameters from Planck PR4 with CamSpec}",
    doi = "10.1093/mnras/stac2744",
    journal = "Mon. Not. Roy. Astron. Soc.",
    volume = "517",
    number = "3",
    pages = "4620-4636",
    year = "2022"
}

@article{Efstathiou:2019mdh,
    author = "Efstathiou, George and Gratton, Steven",
    title = "{A Detailed Description of the CamSpec Likelihood Pipeline and a Reanalysis of the Planck High Frequency Maps}",
    eprint = "1910.00483",
    archivePrefix = "arXiv",
    primaryClass = "astro-ph.CO",
    doi = "10.21105/astro.1910.00483",
    month = "10",
    year = "2019"
}

@article{silva2012,
    title={Thermodynamics and dark energy},
    author={Silva, R and Gonçalves, R S and Alcaniz, J S and Silva, H H B},
    journal={Astron. Astrophys.},
    volume={537},
    pages={A11},
    year={2012},
    publisher={EDP Sciences},
    DOI={https://doi.org/10.1051/0004-6361/201117707}
}

@article{silva2013,
    title={General treatment for dark energy thermodynamics},
    author={Silva, H H B and Silva, R and Gonçalves, R S and {Z-H. Zhu} and Alcaniz, J S},
    journal={Phys. Rev. D},
    volume={88},
    pages={127302},
    year={2013},
    publisher={APS},
    DOI={https://doi.org/10.1103/PhysRevD.88.127302}
}

@article{lima2008,
    title={Chemical potential and the nature of dark energy: The case of a phantom field},
    author={Lima, J A S and Pereira, S H},
    journal={Phys. Rev. D},
    volume={78},
    pages={083504},
    year={2008},
    publisher={APS},
    DOI={https://doi.org/10.1103/PhysRevD.78.083504}
}

@article{pereira,
    title={On phantom thermodynamics},
    author={Pereira, S H and Lima, J A S},
    journal={Phys. Lett. B},
    volume={669},
    pages={266-270},
    year={2008},
    publisher={Elsevier},
    DOI={https://doi.org/10.1016/j.physletb.2008.10.006}
}

@article{calvao,
    title={On the thermodynamics of matter creation in cosmology},
    author={Calvão, M O and Lima, J A S and Waga, I},
    journal={Phys. Lett. A},
    volume={162},
    pages={223-226},
    year={1992},
    publisher={Elsevier},
    DOI={https://doi.org/10.1016/0375-9601(92)90437-Q}
}

@article{lima1992,
    title={On the equivalence of bulk viscosity and matter creation},
    author={Lima, J A S and Germano, A S M},
    journal={Phys. Lett. A},
    volume={170},
    pages={373-378},
    year={1992},
    publisher={Elsevier},
    DOI={https://doi.org/10.1016/0375-9601(92)90890-X}
}

@article{zimdahl1993,
    title={Cosmology with adiabatic matter creation},
    author={Zimdahl, W and Pavón, D},
    journal={Phys. Lett. A},
    volume={176},
    pages={57-61},
    year={1993},
    publisher={Elsevier},
    DOI={https://doi.org/10.1016/0375-9601(93)90316-R}
}

@article{lima1996,
    title={{FRW-type cosmologies with adiabatic matter creation}},
    author={Lima, J A S and Germano, A S M and Abramo, L R W},
    journal={Phys. Rev. D},
    volume={53},
    pages={4287},
    year={1996},
    publisher={APS},
    DOI={https://doi.org/10.1103/PhysRevD.53.4287}
}

@article{lima1999.1,
    author={J A S Lima and J S Alcaniz},
    title={{Flat FRW Cosmologies with adiabatic matter creation: Kinematic tests}},
    eprint={astro-ph/9902337},
    archivePrefix={arXiv},
    primaryClass={astro-ph.CO},
    DOI={10.48550/arXiv.astro-ph/9902337},
    year={1999}
}

@article{lima1999.2,
    title={{Closed and open FRW cosmologies with matter creation: Kinematic tests}}, 
    author={J A S Lima and J S Alcaniz},
    year={1999},
    eprint={astro-ph/9906410},
    archivePrefix={arXiv},
    primaryClass={astro-ph.CO},
    DOI={10.48550/arXiv.astro-ph/9906410}
}

@article{zimdahl1996.2,
    title={Bulk viscous cosmology},
    author={Zimdahl, W},
    journal={Phys. Rev. D},
    volume={53},
    pages={5483},
    year={1996},
    publisher={APS},
    DOI={https://doi.org/10.1103/PhysRevD.53.5483}
}

@article{costanetto,
    title = {A thermodynamic model for dark energy including particle creation or destruction processes},
    author = {Costa Netto, J M and H H B Silva},
    journal = {Chin. J. Phys.},
    volume = {94},
    pages = {684-689},
    year = {2025},
    publisher={Elsevier},
    DOI={https://doi.org/10.1016/j.cjph.2025.02.007}
}

@article{chevallier,
  title={Accelerating universes with scaling dark matter},
  author={Chevallier, M and Polarski, D},
  journal={Int. J. Mod. Phys. D},
  volume={10},
  number={02},
  pages={213-223},
  year={2001},
  publisher={World Scientific},
  DOI={https://doi.org/10.1142/S0218271801000822}
}

@article{linder,
  title={Exploring the expansion history of the universe},
  author={Linder, E V},
  journal={Phys. Rev. Lett.},
  volume={90},
  number={9},
  pages={091301},
  year={2003},
  publisher={APS},
  DOI={https://doi.org/10.1103/PhysRevLett.90.091301}
}

@article{barboza,
  title={A parametric model for dark energy},
  author={{Barboza Jr.}, E M and Alcaniz, J S},
  journal={Phys. Lett. B},
  volume={666},
  number={5},
  pages={415-419},
  year={2008},
  publisher={Elsevier},
  DOI={https://doi.org/10.1016/j.physletb.2008.08.012}
}

@article{planck2018.1,
    title={{Planck 2018 results-VI. Cosmological parameters}},
    author={Aghanim, N and others},
    journal={Astron. Astrophys.},
    volume={641},
    pages={A6},
    year={2020},
    publisher={EDP sciences},
    DOI={https://doi.org/10.1051/0004-6361/201833910}
}

@article{planck2018.2,
  title={{Planck 2018 results-I. Overview and the cosmological legacy of Planck}},
  author={Aghanim, N and others},
  journal={Astron. Astrophys.},
  volume={641},
  pages={A1},
  year={2020},
  publisher={EDP sciences},
  DOI={https://doi.org/10.1051/0004-6361/201833880}
}

@article{pantheon2022.1,
  title={The Pantheon+ analysis: the full data set and light-curve release},
  author={Scolnic, D and others},
  journal={Astrophys. J.},
  volume={938},
  number={2},
  pages={113},
  year={2022},
  publisher={IOP Publishing},
  DOI={10.3847/1538-4357/ac8b7a}
}

@article{pantheon2022.2,
  title={The Pantheon+ analysis: cosmological constraints},
  author={Brout, D and others},
  journal={Astrophys. J.},
  volume={938},
  number={2},
  pages={110},
  year={2022},
  publisher={IOP Publishing},
  DOI={10.3847/1538-4357/ac8e04}
}

@article{union3,
  title={{Union through UNITY: Cosmology with 2000 SNe Using a Unified Bayesian Framework}},
  author={Rubin, D and others},
  journal={Astrophys. J.},
  volume={986},
  number={2},
  pages={231},
  year={2025},
  publisher={IOP Publishing},
  DOI={10.3847/1538-4357/adc0a5}
}

@article{des,
  title={{The dark energy survey: Cosmology results with~ 1500 new high-redshift type Ia supernovae using the full 5 yr data set}},
  author={Abbott, T M C and others},
  journal={Astrophys. J.},
  volume={973},
  number={1},
  pages={L14},
  year={2024},
  publisher={IOP Publishing},
  DOI={10.3847/2041-8213/ad6f9f}
}

@book{weinberg1972,
    title={Gravitation and cosmology: principles and applications of the general theory of relativity},
    author={Weinberg, S},
    year={1972},
    publisher={John Wiley and Sons},
    address={New York},
    edition={1}
}

@book{weinberg2008,
    title={Cosmology},
    author={Weinberg, S},
    year={2008},
    publisher={OUP Oxford},
    address={New York},
    edition={1}
}

@article{silva2002,
    title={Temperature evolution law of imperfect relativistic fluids},
    author={Silva, R and Lima, J A S and Calvão, M O},
    journal={Gen. Relativ. Gravit.},
    volume={34},
    pages={865-875},
    year={2002},
    publisher={Springer},
    DOI={https://doi.org/10.1023/A:1016317914912}
}

@article{eckart,
    title={The thermodynamics of irreversible processes III. Relativistic theory of the simple fluid},
    author={Eckart, C},
    journal={Phys. Rev.},
    volume={58},
    number={10},
    pages={919},
    year={1940},
    publisher={APS},
    DOI={https://doi.org/10.1103/PhysRev.58.919}
}

@book{landau,
  title={Fluid mechanics: Volume 6},
  author={Landau, L D and Lifshitz, E M},
  year={1987},
  publisher={Pergamon Press},
  address={Oxford},
}

@article{gunzig,
    title={Entropy and cosmology},
    author={Gunzig, E and Geheniau, J and Prigogine, I},
    journal={Nature},
    volume={330},
    pages={621-624},
    year={1987},
    publisher={Nature Publishing Group UK London},
    DOI={https://doi.org/10.1038/330621a0}
}

@article{prigogine1988,
    title={Thermodynamics of cosmological matter creation},
    author={Prigogine, I and Geheniau, J and Gunzig, E and Nardone, P},
    journal={Proc. Natl. Acad. Sci. USA},
    volume={85},
    pages={7428-7432},
    year={1988},
    publisher={National Acad Sciences},
    DOI={https://doi.org/10.1073/pnas.85.20.7428}
}

@article{prigogine1989,
    title={Thermodynamics and cosmology},
    author={Prigogine, I},
    journal={Int. J. Theor. Phys.},
    volume={28},
    pages={927-933},
    year={1989},
    publisher={Springer},
    DOI={https://doi.org/10.1007/BF00670337}
}

@article{lima2004,
    title={Thermodynamics, spectral distribution and the nature of dark energy},
    author={Lima, José A S and Alcaniz, Jailson S},
    journal={Phys. Lett. B},
    volume={600},
    pages={191-196},
    year={2004},
    publisher={Elsevier}
}

@article{Gonzalez:2018rop,
    author = "Gonzalez, J. E. and Silva, H. H. B. and Silva, R. and Alcaniz, J. S.",
    title = "{Physical constraints on interacting dark energy models}",
    journal = "Eur. Phys. J. C",
    volume = "78",
    number = "9",
    pages = "730",
    year = "2018",
    DOI={https://doi.org/10.1140/epjc/s10052-018-6212-3}
}

@article{GetDist,
    author = "Lewis, Antony",
    title = "{GetDist: a Python package for analysing Monte Carlo samples}",
    doi = "10.1088/1475-7516/2025/08/025",
    journal = "J. Cosmol. Astropart. Phys.",
    volume = "2025",
    pages = "025",
    year = "2025"
}

@article{CLASS,
    author = "Blas, Diego and Lesgourgues, Julien and Tram, Thomas",
    title = "{The Cosmic Linear Anisotropy Solving System (CLASS) II: Approximation schemes}",
    doi = "10.1088/1475-7516/2011/07/034",
    journal = "J. Cosmol. Astropart. Phys.",
    volume = "2011",
    pages = "034",
    year = "2011"
}

@article{carron,
  title={CMB lensing from Planck PR4 maps},
  author={Carron, Julien and Mirmelstein, Mark and Lewis, Antony},
  journal={J. Cosmol. Astropart. Phys.},
  volume={2022},
  number={9},
  pages={039},
  year={2022},
  publisher={IOP Publishing},
  DOI={10.1088/1475-7516/2022/09/039}
}

@article{COBAYA,
    author = "Torrado, Jesus and Lewis, Antony",
    title = "{Cobaya: Code for Bayesian Analysis of hierarchical physical models}",
    doi = "10.1088/1475-7516/2021/05/057",
    journal = "J. Cosmol. Astropart. Phys.",
    volume = "2021",
    pages = "057",
    year = "2021"
}

@article{ratra,
  title={Cosmological consequences of a rolling homogeneous scalar field},
  author={Ratra, Bharat and Peebles, Philip JE},
  journal={Phys. Rev. D},
  volume={37},
  number={12},
  pages={3406},
  year={1988},
  publisher={APS},
  DOI={https://doi.org/10.1103/PhysRevD.37.3406}
}

@article{caldwell,
  title={Cosmological imprint of an energy component with general equation of state},
  author={Caldwell, Robert R and Dave, Rahul and Steinhardt, Paul J},
  journal={Phys. Rev. Lett.},
  volume={80},
  number={8},
  pages={1582},
  year={1998},
  publisher={APS},
  DOI={https://doi.org/10.1103/PhysRevLett.80.1582}
}

@article{zlatev,
  title={Quintessence, cosmic coincidence, and the cosmological constant},
  author={Zlatev, Ivaylo and Wang, Limin and Steinhardt, Paul J},
  journal={Phys. Rev. Lett.},
  volume={82},
  number={5},
  pages={896},
  year={1999},
  publisher={APS},
  DOI={https://doi.org/10.1103/PhysRevLett.82.896}
}

@article{peebles,
  title={The cosmological constant and dark energy},
  author={Peebles, P James E and Ratra, Bharat},
  journal={Rev. Mod. Phys.},
  volume={75},
  number={2},
  pages={559},
  year={2003},
  publisher={APS},
  DOI={https://doi.org/10.1103/RevModPhys.75.559}
}
\end{document}